\documentclass[prl,aps,showpacs,twocolumn,preprintnumbers,
amsmath,amssymb,superscriptaddress,longbibliography]{revtex4-2}
\usepackage[english]{babel}

\usepackage{amsmath,amssymb,amsfonts}
\usepackage{graphicx}
\usepackage[colorlinks=True,linkcolor=red,citecolor=blue,urlcolor=blue]{hyperref}
\usepackage{bookmark}
\usepackage[dvipsnames]{xcolor}
\usepackage{chngcntr}
\usepackage{braket}
\usepackage{nicefrac}
\usepackage{bm}
\usepackage{bbm}
\usepackage{braket}
\usepackage{slashed}
\usepackage[normalem]{ulem}
\usepackage{array}
\newcolumntype{C}{>{$}c<{$}}
\usepackage{chemmacros}

\renewcommand{\Im}{\mathop{\mathrm{Im}}}

\newcommand{\smeq}{\! = \!}

\newcommand\scalemath[2]{\scalebox{#1}{\mbox{\ensuremath{\displaystyle #2}}}}
\allowdisplaybreaks

\def\ket#1{|#1\rangle }
\def\bra#1{\langle #1 |}

\usepackage[sectionbib]{bibunits}
\begin{document}
\defaultbibliographystyle{apsrev4-1} 
\defaultbibliography{Biblio} 

\title{
A local quantized marker for topological magnons from circular dichroism}

\author{Baptiste Bermond}
\email{baptiste.bermond@lkb.ens.fr}
\affiliation{Laboratoire Kastler Brossel, Coll\`ege de France, CNRS, ENS-Universit\'e PSL,
Sorbonne Universit\'e, 11 Place Marcelin Berthelot, 75005 Paris, France
}
\author{Ana\"is Defossez}
\affiliation{Laboratoire Kastler Brossel, Coll\`ege de France, CNRS, ENS-Universit\'e PSL,
Sorbonne Universit\'e, 11 Place Marcelin Berthelot, 75005 Paris, France
}
\affiliation{International Solvay Institutes, 1050 Brussels, Belgium}
\affiliation{Center for Nonlinear Phenomena and Complex Systems, Universit\'e Libre de Bruxelles, CP 231, Campus Plaine, B-1050 Brussels, Belgium}

\author{Gregor Jotzu}
\affiliation{Dynamic Quantum Materials Laboratory, Institute of Materials, EPFL, 1015 Lausanne, Switzerland}

\author{Nathan Goldman}
\email{nathan.goldman@lkb.ens.fr}
\affiliation{Laboratoire Kastler Brossel, Coll\`ege de France, CNRS, ENS-Universit\'e PSL,
Sorbonne Universit\'e, 11 Place Marcelin Berthelot, 75005 Paris, France
}
\affiliation{International Solvay Institutes, 1050 Brussels, Belgium}
\affiliation{Center for Nonlinear Phenomena and Complex Systems, Universit\'e Libre de Bruxelles, CP 231, Campus Plaine, B-1050 Brussels, Belgium}

\begin{abstract}
 The low-energy excitations of a spin system can display Bloch bands with non-trivial topological properties. While topological magnons can be identified through the detection of chiral propagating modes at the sample's edge, an intriguing approach would be to directly probe their topological nature via localized measurements deep within the bulk. In this work, we introduce a quantized topological marker suitable for topological spin systems,  which can be experimentally accessed by combining a local driven-dissipative preparation scheme with a circular-dichroic measurement. Demonstrated on a $2$D ferromagnetic Heisenberg spin system incorporating Dzyaloshinskii–Moriya interactions, this method effectively maps a local Chern marker with single-site resolution while inherently accounting for magnon losses. Our work offers a general strategy to access local topological markers in bosonic settings within a driven-dissipative framework.
\end{abstract}
\date{\today}
\maketitle 
\textit{Introduction ---} Certain materials can exhibit low-energy bosonic excitations with topological band properties~\cite{cao2015magnon,owerre2016first,mcclarty2022topological,subramanian2024topological,zhuo2025topological}. This is the case of topological magnons, which can be found in materials such as \iupac{Cu[1,3-benzene|di|carboxylate]}~\cite{chisnell2015topological,hirschberger2015thermal}, \ch{Na3Cu2SbO6}~\cite{miura2006spin}, \iupac{\chembeta-\ch{Cu2V2O7}} ~\cite{tsirlin2010beta}, \ch{Mn5Ge3}~\cite{dos2023topological} and \ch{CrI3}~\cite{chen2018topological}. A direct manifestation of these topological properties is provided by the chiral edge excitations, which can be probed spectroscopically~\cite{malz2019topological,feldmeier2020local}. In this context, a relevant question concerns whether one could access a quantized topological response, deep in the bulk of the system, which would directly reflect the topological nature of the magnonic Bloch bands. Besides, edge magnonic excitations have been shown to be affected by losses and dissipation~\cite{habel2024breakdown}. It would therefore be appealing to develop a realistic scheme to access topological responses of magnonic excitations that would remain immune to such losses, i.e.~within an engineered driven-dissipative scenario.

More generally, the identification of local markers of topology has been an important quest in the realm of topological quantum matter. Prime instances are provided by the so-called Bianco-Resta marker~\cite{bianco2011mapping}, and also by the local Streda response~\cite{Resta_Streda,umucalilar2008trapped,Cecile_Streda_FCI,wang2024cold,markov2024locality}, which allows for the local determination of the (many-body) Chern number deep in the bulk of a Chern-insulating system. However, these markers have been developed and applied chiefly to fermionic and/or charged settings, and their extension to magnons -- charge-neutral bosonic excitations -- is far from immediate:~while topological Bloch bands are naturally filled in fermionic settings simply by adjusting the chemical potential, the ground state of a bosonic system is typically an empty vacuum; moreover, magnons do not couple to the traditional experimental probes of topological matter. Altogether, any strategy aimed at measuring the local topological properties of magnonic excitations must be accompanied by a well-defined preparation and probing protocol.

\begin{figure}[h!]
    \centering
    \includegraphics[width=\linewidth]{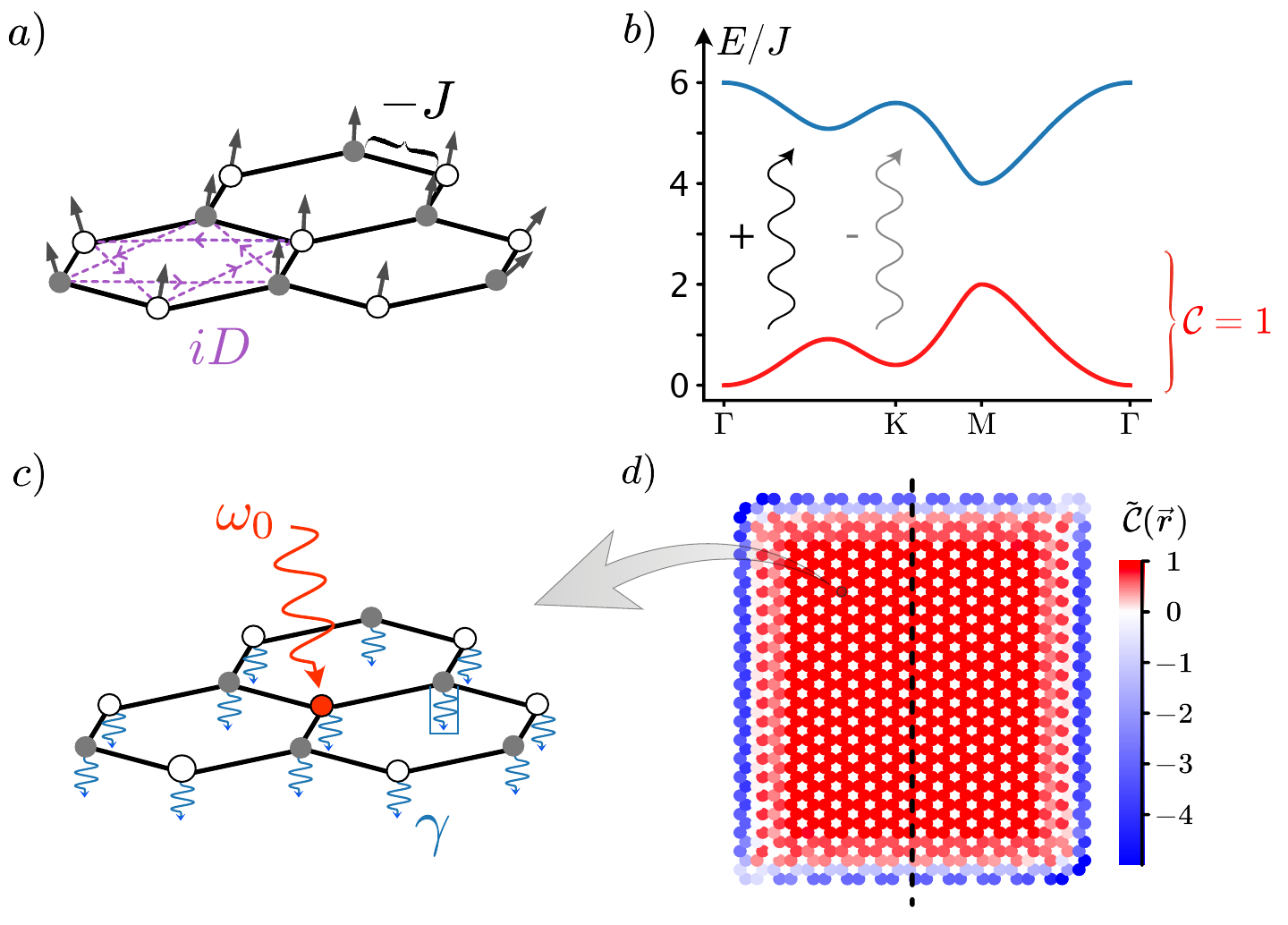}
    \caption{\textit{Local Chern marker measurement in a topological magnon system.} (a) Illustration of the $2$D spin model in Eq.~\eqref{eq:spin_ham}, with nearest-neighbor ferromagnetic Heisenberg interactions $J$ (black), and next-nearest-neighbor DM interactions $D$ (blue). (b) The corresponding energy spectrum displays two bands, with respective Chern numbers $\mathcal{C}\!=\!\pm1$; here $J\!=\!2D>0$. (c) Sketch of the preparation scheme:~a localized spin excitation is created using a local pump field at frequency $\omega_0$, in the presence of a homogeneous loss rate $\gamma$. The resulting steady-state uniformly occupies the low-energy magnonic band. A circular perturbation (not shown) is then applied to drive transitions to the upper band [see panel (b)]. (d) The Chern marker, as extracted from this local circular-dichroic measurement, is shown for a $20 \times 20$ spin system.
    }
    \label{fig:DMI}
\end{figure}

In this Letter, we introduce a local topological marker that is experimentally accessible in a broad range of bosonic settings hosting Chern bands, such as topological magnons. Our construction consists of three steps: (i)~the preparation of a localized (single-site) excitation, deep in the bulk, which entirely projects onto a single topological band; (ii)~the application of a circular drive, which acts globally on the system; (iii)~the monitoring of excitation rates (or power absorbed) upon the action of the circular drive, for the two opposite orientations of the drive~\cite{tran2017probing,repellin2019detecting,asteria2019measuring,goldman2024relating,10.21468/SciPostPhysCore.6.3.059}. This local circular-dichroic measurement eventually provides the desired topological marker, revealing the Chern number of the underlying band at the single-site level. This protocol is further generalized into a space- and energy-resolved approach, enabling robust topological characterization in systems with highly dispersive bands and small bandgaps. Importantly, the preparation stage can be achieved through a driven-dissipative approach, which potentially guarantees the robustness of this approach to inevitable losses. While this construction generally applies to bosonic Chern-band systems, we illustrate it on the concrete case of magnonic excitations in a $2$D collinear ferromagnetic Heisenberg spin system~\cite{owerre2016first}.\\

\textit{Model and spin-wave theory ---} For concreteness, let us consider a prototypical model of topological magnon systems~\cite{owerre2016first,kim2016realization}. This model consists of a ferromagnetic material with spin-orbit coupling, whose localized spins are located on a honeycomb lattice. The Hamiltonian is expressed as [Fig.~\ref{fig:DMI}(a)]:
\begin{equation}
    \label{eq:spin_ham}
    \hat{\mathcal{H}}=-J\sum_{\langle ij\rangle}\Vec{s}_i\cdot\Vec{s}_j+\sum_{\langle\langle ij\rangle\rangle}\Vec{D}_{ij}\cdot\left(\Vec{s}_i\wedge\Vec{s}_j\right)\,,
\end{equation}
where the spin operators $\Vec{s}_i$ describe spin-degrees of
freedom of magnitude $S$. The first term of Hamiltonian~\eqref{eq:spin_ham} is a nearest-neighbor Heisenberg interaction, and we set $J\!>\!0$ to stabilize the ferromagnetic order. The second term is the Dzyaloshinskii–Moriya (DM) interaction between the next-nearest neighboring spins. Since the honeycomb plane is a mirror plane of the system,
the DM vector does not have any in-plane component~\cite{moriya1960anisotropic,gao2019thermal}, i.e. $\Vec{D}_{ij}\!=\!D_{ij}\Vec{e}_z$,  with $\Vec{e}_z$ the unit vector in the z-direction. Considering a uniform system, we set $D_{ij}\!=\!-D_{ji}\!=\!D\eta_{ij}$, where the sign $\eta_{ij}\!=\!\pm1$ depends on the orientation of the two next-nearest spins [Fig.~\ref{fig:DMI}(a)].
Interestingly, this Hamiltonian commutes with the z-component of the total spin $\hat{S}_z=\sum_i \hat{s}_i^z$. As a consequence, the total magnetization in the z-direction $S_z=\langle\Hat{S}_z\rangle$ is a conserved quantity, allowing one to study sectors of different magnetizations independently.

Provided that $D$ is small enough ($D\lesssim0.7$) the ground state is ferromagnetic and the linear spin wave Hamiltonian reads [End Matter]
\begin{equation}
    \label{eq:LSW_H}
    \begin{aligned}
        \hat{\mathcal{H}}
        =&-N_lJS^2-JS\sum_{\langle ij\rangle}\left(\hat{a}^\dagger_j\hat{a}_i+\hat{a}^\dagger_i\hat{a}_j\right)\\&+JS\sum_iz_i\hat{a}^\dagger_i\hat{a}_i+iDS\sum_{\langle\langle ij\rangle\rangle}\eta_{ij}\left(\hat{a}^\dagger_j\hat{a}_i-\hat{a}^\dagger_i\hat{a}_j\right),
    \end{aligned}
\end{equation}
where $\hat a_i^\dagger$($\hat a_i$) creates (anihilates) a magnon at site $i$, with $N_l=\sum_iz_i$ the total number of links, and $z_i$ the number of nearest neighbors of site $i$, i.e., $z_i = 3$ in the bulk, and $z_i\le 3$ on the edges. This Hamiltonian~\eqref{eq:LSW_H} describes a bosonic analog of the two-band Haldane model~\cite{haldane1988model}, with a staggered $\pi/2$ phase and a topological band-gap of size $6\sqrt{3}DS$, whose spectrum is represented in Fig.~\ref{fig:DMI}(b); see~\cite{SuppMat} for an application to a three-band model.



\begin{figure}
    \centering
    \includegraphics[width=\linewidth]{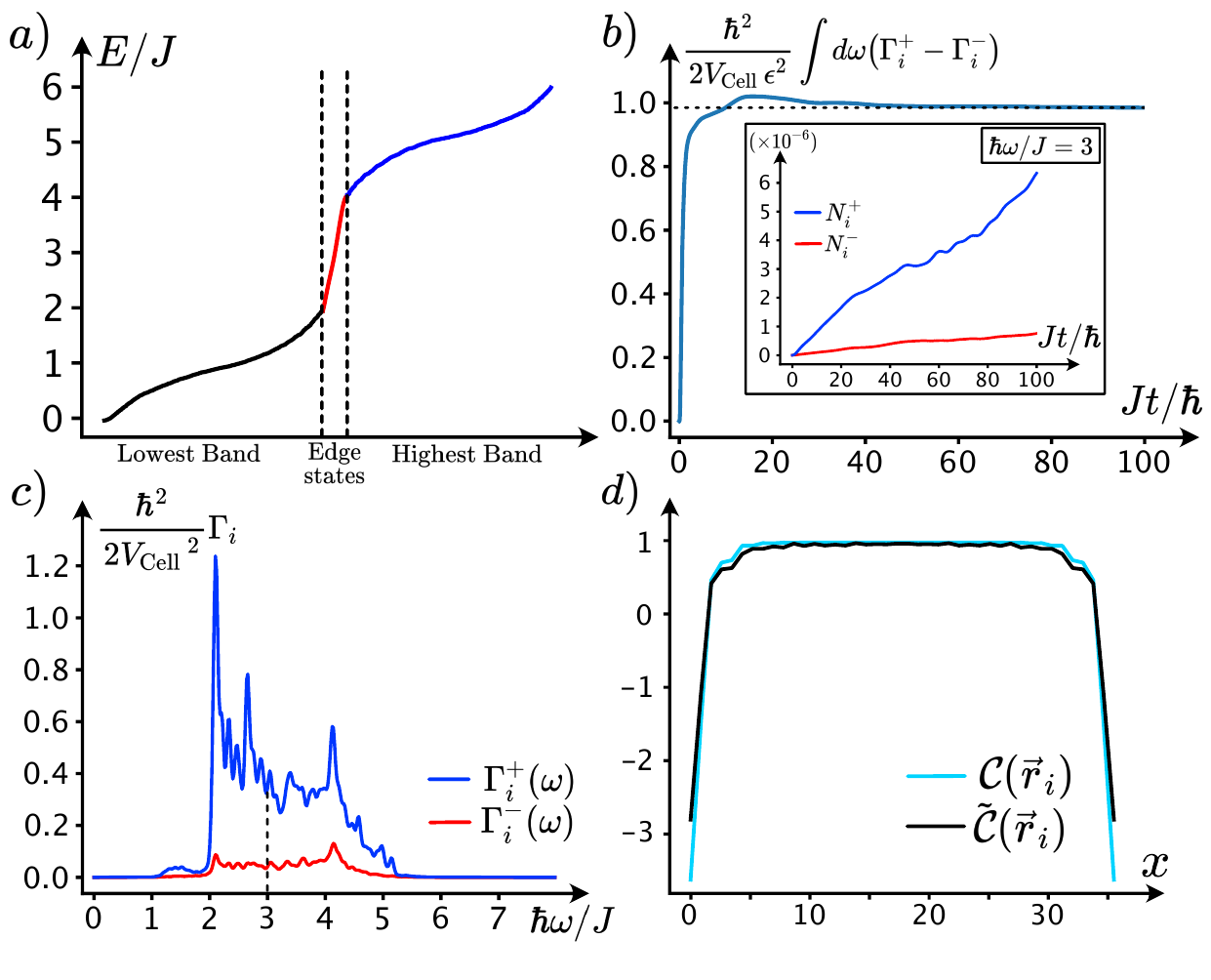}
    \caption{\textit{Dichroism measurement for the $S\!\smeq\!1/2$ bosonized Hamiltonian~\eqref{eq:LSW_H}.} (a) Single-magnon spectrum for a $20\times20$ lattice and $D \!\smeq\!J/2$:~the states in the lowest (highest) bulk band are represented in black (blue), while the edge states are shown in red. (b) Time evolution of the integrated differential rate [Eq.~\eqref{eq:rate_diff}] as a function of time, for a specific initial state located in the bulk; the driving amplitude is $\epsilon\!\smeq \!2.10^{-4}J/a$. In the long-time limit, this quantity converges to a quantized value, defining a local Chern marker. The inset shows the time evolution of the excited fractions $N_{i}^{\pm}(\omega,t)$ for $\hbar\omega/J\!=\!3$, corresponding to the dashed line in panel (c). (c) The late-time excitation rates $\Gamma_{i}^{\pm}(\omega,Jt/\hbar=100)$ as a function of $\hbar \omega /J$. (d) The Chern marker $\widetilde{\mathcal{C}}(\Vec{r}_i)$ (black) is compared to the Bianco-Resta marker $\mathcal{C}(\Vec{r}_i)$ (blue); both markers are evaluated along the dashed line depicted in Fig.~\ref{fig:DMI}(d). The local Chern markers exhibit a quantized value ($\mathcal{C}=\widetilde{\mathcal{C}}=1$) in the bulk, while they take large negative values near the edges.}
    \label{fig:bosonized_results}
\end{figure}
\textit{A practical Chern marker for topological magnons ---} Topological magnon properties can be captured at the single-magnon level. As an initial state, we thus consider a single-magnon state that uniformly occupies the lowest band of the spectrum and that is localized on a single site of the lattice. Theoretically, such an initial state $\left|\downarrow_i\right\rangle$ can be computed as the projection onto the lowest band of the single-magnon excitation $\left|i\right\rangle=\hat{a}^\dagger_i\left|0\right\rangle$, with $\left|0\right\rangle$ the vacuum state defined by $\hat{a}_i\left|0\right\rangle=0$ for all sites $i$:
\begin{equation}
\label{eq:initial_boson}
    \ket{\downarrow_i}=\frac{\sum_{g\in LB}\ket{g}\braket{g|i}}{\sqrt{\sum_{g\in LB} \left|\braket{g|i}\right|^2}}\,\quad\,, 
\end{equation}
where $\sum_{g\in LB}$ denotes a sum over all the eigenstates $|g\rangle$ belonging to the lowest energy band [Fig.~\ref{fig:bosonized_results}(a)].\\
We now analyze the response of this localized state to a global circular drive. We introduce this chiral perturbation in the form of a time-periodic potential
\begin{equation}
\label{eq:chiral_pert}
    \delta\hat{\mathcal{H}}_\pm(t)=2\epsilon\left[\hat{x}\cos\left(\omega t\right)\pm\hat{y}\sin\left(\omega t\right)\right]\,,
\end{equation}
where $\hat{x}/\hat{y}$ denote the position operators defined as $\hat{x}\!=\!\sum_jx_j\hat{a}^\dagger_j\hat{a}_j$, with $x_i$ the x coordinate of the site $i$; $\omega$ is the drive frequency, $\epsilon$ is the amplitude, and $\pm$ denotes the two drive orientations. Assuming a chiral drive of small amplitude, we evaluate $N_i^\pm(\omega,t)$ the excited fraction in the higher band; see the inset of Fig.~\ref{fig:bosonized_results}(b) for illustrative time evolutions. The circular dichroic signal is then obtained by comparing the long-time chiral excitation rates,
\begin{equation}
    \Delta\Gamma_i(\omega,t)=\frac{N_i^+(\omega,t)-N_i^-(\omega,t)}{2t}\, .
\end{equation}
In practice, and as we further discuss below, the chiral excitation rates $\Gamma_i^\pm(\omega,t)\!\equiv\! N_i^{\pm}(\omega,t)/t$ can be measured by monitoring the power absorbed:~$P_i^\pm(\omega,t)=\hbar\omega\Gamma_i^\pm(\omega,t)$. Figure~\ref{fig:bosonized_results}(c) represents an illustrative chiral absorption spectrum $\Gamma_i^\pm(\omega,t)$. Inspired by Refs.~\cite{tran2017probing,asteria2019measuring}, we consider performing a set of experiments for various frequencies of the drive $\omega$, and integrating the circular dichroic signal over the frequencies, which, at the first order in time-dependent perturbation theory, reads~\cite{SuppMat}
\begin{align}
\Delta\Gamma_i(t)&=\int_0^{+\infty}\textrm{d}\omega\Delta\Gamma_i(\omega,t)=\int_0^{+\infty}\textrm{d}\omega\frac{\Delta P_i(\omega,t)}{\hbar\omega} \label{eq:rate_diff} \\
&\underset{t\to\infty}{=}-\frac{4\pi \epsilon^2}{\hbar^2}{\rm Im}\left(\sum_{g\in LB}\braket{ \downarrow_i|g}\langle g|\hat{x}\hat{Q}\hat{y}|g\rangle\braket{ g|\downarrow_i}\right)\,, \notag 
\end{align}
where $\hat{Q}$ is the projector onto the states belonging to the higher bands [Fig.~\ref{fig:bosonized_results}(a)]. We point out that this approach applies to multi-band systems; see Ref.~\cite{SuppMat} for an application to a three-band model. Figure~\ref{fig:bosonized_results}(b) displays an illustrative time evolution of the integrated circular dichroic signal $\Delta\Gamma_i(t)$. From its late time limit, we finally define a local marker $\widetilde{\mathcal{C}}(\Vec{r}_i)$ as 
\begin{align}
    \widetilde{\mathcal{C}}(\Vec{r}_i)&\equiv\underset{t\to \infty}{\textrm{lim}}\left(\frac{\hbar^2}{V_{{\rm cell}}\epsilon^2}\Delta\Gamma_i (t)\right) \label{eq:Chern_Mark_us} \\
    &=-\frac{4\pi}{V_{{\rm cell}}}{\rm Im}\left(\sum_{g\in LB}\braket{ \downarrow_i|g}\bra{g}\hat{x}\hat{Q}\hat{y}\ket{g}\braket{g|\downarrow_i}\right)\,, \notag 
\end{align}
where $V_{\rm cell}$ denotes the area of the unit cell. Notably, this marker closely resembles the local Chern marker introduced by Bianco and Resta~\cite{bianco2011mapping}; see End Matter for further discussion. As shown in Figs.~\ref{fig:DMI}(d) and \ref{fig:bosonized_results}(d), $\widetilde{\mathcal{C}}(\Vec{r}_i)$ displays behavior consistent with a local Chern marker:~it is quantized to 1 in the bulk and large and negative near the edges. Finite-size scaling at the sample center confirms exponential convergence of both $\widetilde{\mathcal{C}}(\Vec{r})$ and its finite-time version $\widetilde{\mathcal{C}}(\Vec{r},t)$ to the quantized value; see Ref.~\cite{SuppMat}.

\textit{Local Chern marker in spin models ---} By reversing the bosonization procedure, one can express the single-magnon initial state $\ket{\downarrow_i}$ as well as the perturbation in Eq.~\eqref{eq:chiral_pert} in the spin language; this mapping allows one to describe a practical procedure to access the local Chern marker in spin systems through circular dichroism. In the spin language, the equivalent of single-boson states corresponds to states of the magnetization sector $S_z=NS-1$ with $N$ the number of sites and $S$ the spin amplitude, such that the equivalent of $|\downarrow_i\rangle$ is given by
    \begin{equation}
\label{eq:initial_spin}
    \ket{\downarrow_i^s}=\frac{\sum_{g\in LB}\ket{g}\braket{g|i}}{\sqrt{\sum_{g\in LB} \left|\braket{g|i}\right|^2}}\,,\quad\,\textrm{with} \left|i\right\rangle=\hat{s}^-_i\left|0\right\rangle\,,
\end{equation}
with $\hat{s}^\pm_j=\hat{s}^x_j\pm i\hat{s}^y_j$, $|0\rangle$ the ground state defined by $\hat{s}_i^+\left|0\right\rangle=0$ for all sites $i$, and where $|g\rangle$ denotes the eigenstates belonging to the lowest magnon energy band; see Ref.~\cite{SuppMat} for a representation (amplitude and phase) of the corresponding initial state. While this construction assumes zero temperature, we expect the results to remain applicable provided the temperature is small compared to the spin gap, which is estimated to be on the order of a kelvin in the relevant material \ch{CrI3}~\cite{chen2018topological,chen2020magnetic}. As the probe power can be arbitrarily weak, heating rates can be reduced to a level appropriate for a dilution cryostat.

Applying the spin–boson mapping to circular driving, we consider a chiral perturbation of the form~\eqref{eq:chiral_pert}, where $\hat{x}$ and $\hat{y}$ now denote the spin position operators defined via the Holstein–Primakoff transformation [End Matter]:
\begin{equation}   
\delta\hat{\mathcal{H}}_\pm(t)=- 2\epsilon \sum_j \left[x_j \cos\left(\omega t\right)\pm y_j\sin\left(\omega t\right)\right] \hat{s}_j^z . \label{x_spin}
\end{equation}
In practice, the chiral perturbation in Eq.~\eqref{x_spin} can be implemented as (anti-)clockwise rotating magnetic field gradients generating a spatially dependent Zeeman term, which can be produced using focused azimuthally polarized vortex beams \cite{richards1959electromagnetic, khonina2016ultrafast} or off-centered phase-locked grazing incidence light. The dichroic measurement could then be performed by extracting the absorbed power $P_i^\pm(\omega,t)$ through analysis of the transmitted and reflected light across probing frequencies.


The numerical study of our local circular-dichroic scheme can be performed at the level of the spin description. We verified that the results are strictly identical to those presented in Fig.~\ref{fig:bosonized_results} for the bosonic model within the present single-magnon framework. \\


\begin{figure}
    \centering
    \includegraphics[width=\linewidth]{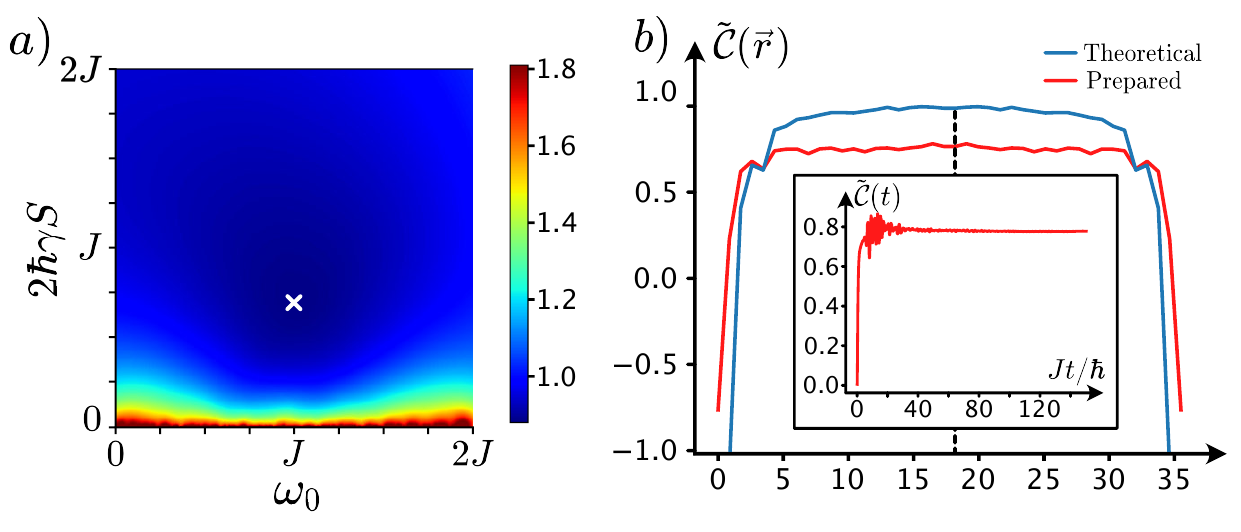}
    
    \caption{\textit{Driven-dissipative preparation scheme and local Chern marker measurement.} (a) Uniformity factor $\mathcal U$ [Eq.~\eqref{eq:unif_factor}] for the steady-state solution of the mean-field equations [Eq.~\eqref{eq:mean_field_large}], as a function of the pump frequency $\omega_0$ and the loss rate $2\hbar \gamma S$. Its minimal value is marked by a white cross, which is located at $\hbar\omega_0\!=\!J$ and $2\hbar\gamma S\!=\!0.695J$. (b) The local Chern marker $\widetilde{\mathcal{C}}(\Vec{r})$, as obtained from a dichroic measurement performed on the steady state, using the optimal pump-loss parameters [panel (a)]. The marker is displayed along the dashed line represented in Fig.~\ref{fig:DMI}(d). At each position, we extracted the long-time limit of the integrated differential rate (inset). The resulting local marker slightly deviates from a quantized value ($\widetilde{\mathcal{C}}\!\sim\!0.8$); the marker obtained from the ideal initial state is shown in blue for comparison.}
    \label{fig:preparation}
\end{figure}

\textit{Initial state preparation ---} We now discuss how the single-magnon initial states $\vert\!\downarrow_i^s \rangle$ introduced above  [Eqs.~\eqref{eq:initial_boson},\eqref{eq:initial_spin}] can be realized in practice, in an experimental context. Historically, it has been proposed that magnons can be controlled by coupling the system to a metallic layer traversed by an electric current, either globally~\cite{berger1996emission,slonczewski1996current,gunnink2024electrical} or at the edges~\cite{lee2023electronic}; see also Ref.~\cite{mei2019topological} for possible preparation schemes in cold atoms.

Here, we explore a possible scheme that combines a local external pump with the global inherent losses of spin systems [Fig.~\ref{fig:DMI}(c)]. On the one hand, the  local pumping is included directly at the Hamiltonian level 
\begin{equation}
        \delta\hat{\mathcal{H}}_{pump}=\sum_j\mu_BB_j\left(e^{i\omega_0t}\hat{s}^+_j+e^{-i\omega_0t}\hat{s}^-_j\right)\,.
    \end{equation}
Such a pumping can be engineered experimentally by coupling the system to an external rotating magnetic field 
    \begin{equation}
        \sum_j\mu_B\Vec{B}_j(t)\cdot\Vec{s}_j\,, \quad \Vec{B}_j=2B_j\begin{pmatrix}
            \text{cos}(\omega_0t)\\\text{sin}(\omega_0t)\\0
        \end{pmatrix},
    \end{equation}
with a localized amplitude $B_j\!=\!\delta_{ij}$, chosen to realize the local excitation $\left|\downarrow_i\right\rangle$. In practice, single-spin excitation can be achieved via scanning-probe-based local magnetic fields~\cite{yang2019tuning}, or mediated by coherently controlled magnetic adatoms or impurities~\cite{chen2023harnessing, phark2023electric}. Finite-width effects are addressed in Ref.~\cite{SuppMat}, where a pumping scheme based on a Gaussian profile is analyzed. 

On the other hand, losses are assumed to be local and uniform over the lattice. These losses, associated with random magnon losses (i.e.~spin-flips $\hat{s}^+_i$) at a rate $\gamma$, are included in the system using a Lindblad master equation formalism~\cite{SuppMat}. Considering the large $S$ limit for illustration purposes, and transforming spins to bosons following the Holstein-Primakoff mapping~\eqref{eq:Holstein}, the equation of motion for $b_i=e^{i\omega_0t}\langle \hat{a}_i\rangle$ takes the form 
\begin{equation}
    \label{eq:mean_field_large}
        \begin{aligned}
            i\partial_tb_i=&-JS\sum_{\langle ij\rangle}\left(b_i-b_j\right) -iDS\sum_{\langle\langle ij\rangle\rangle}\eta_{ij}b_j\\
       &-\sqrt{2S}\mu_BB_i-\left(\omega_0+2i\gamma S\right) b_i\,.
        \end{aligned}
    \end{equation}
Similar equations of motion were explored in the context of driven-dissipative photonic lattices~\cite{ozawa2014anomalous,Ozawa_4D,chenier2024quantized,defossez2026energyresolvedquantumgeometrystvreda}.

Given a pump frequency $\omega_0$ and a loss rate $\gamma$, it is possible to obtain a steady-state of Eq.~\eqref{eq:mean_field_large}, $\partial_tb_i\!=\!0$, $\forall i$.  The preparation protocol can then be optimized by identifying the parameters that yield a steady-state that closely resembles the ``ideal" state $\ket{\downarrow_i^s}$. The distance to this ideal state could be evaluated by computing the fidelity $\vert \braket{\psi(t) |\downarrow_i^s} \vert^2$. Instead, we monitor the uniformity with which the steady-state populates the target band, as captured by the ``uniformity factor"
\begin{equation}
  \mathcal{U} =  \sum_{g\in LB}\left||\langle i_{mf}|g\rangle|^2-1/n_{LB}\right|, 
  \label{eq:unif_factor}
\end{equation}
with $|i_{mf}\rangle $ the normalized solution to the mean-field equations in the large S approximation~\eqref{eq:mean_field_large}, and $n_{LB}=\sum_{g\in LB}1$. Identifying the optimal couple $(\omega_0,\gamma)$ then implies minimizing this uniformity criterion.  

In the ideal scenario where the bandwidth $W$ of the lowest Bloch band is small compared to the bandgap $\Delta$, the optimal preparation scheme consists in choosing $\omega_0$ in the middle of the lowest band, with a loss rate $\gamma$ set such that $W\ll\gamma\ll\Delta$; see Refs.~\cite{ozawa2014anomalous,Ozawa_4D}. In the present case, however, the bands are substantially dispersive ($W/\Delta\!\approx\!1$), and we find that the uniformity is maximized for a pump frequency $\hbar\omega_0\!\approx\!J$ and a loss rate $\hbar\gamma \approx 0.695J$; see Fig.~\ref{fig:preparation}(a).

A realistic local Chern marker measurement for spin systems would proceed by preparing a localized initial state via this optimal driven-dissipative scheme, followed by the dichroic protocol described above. An illustrative example of the integrated signal is presented in Fig.~\ref{fig:preparation}(b). Repeating this locally across all sites maps out the marker, which is homogeneous in the bulk with $\widetilde{\mathcal{C}}(\Vec{r})\!\approx\!0.78$, and decreasing near the edges. We have verified that this measurement systematically improves with system size, e.g.~$\widetilde{\mathcal{C}}(\Vec{r})\!\approx\!0.85$ for a $30\times30$ lattice. Still, inevitable deviations from the ideal state (i.e.~$\mathcal{U}\!>\!0$ and higher-band occupation) introduce a small residual error in the Chern marker as the system approaches the thermodynamic limit.

These residual errors primarily stem from spurious occupation of the higher band away from the flat-band limit ($W/\Delta \approx 0$), leading to increasingly large deviations of the marker from the Chern number as the gap decreases. For instance, in $CrI_3$ with $D \approx 0.15J$, yielding $\Delta/W \approx 0.7$~\cite{chen2018topological}, the protocol is expected to produce a significantly non-quantized marker, making it difficult to discriminate between trivial and topological regimes. In the following, we introduce an energy-resolved measurement scheme that mitigates this limitation and extends the approach to systems with highly dispersive bands and small bandgaps.\\

\begin{figure}
    \centering
    \includegraphics[width=\linewidth]{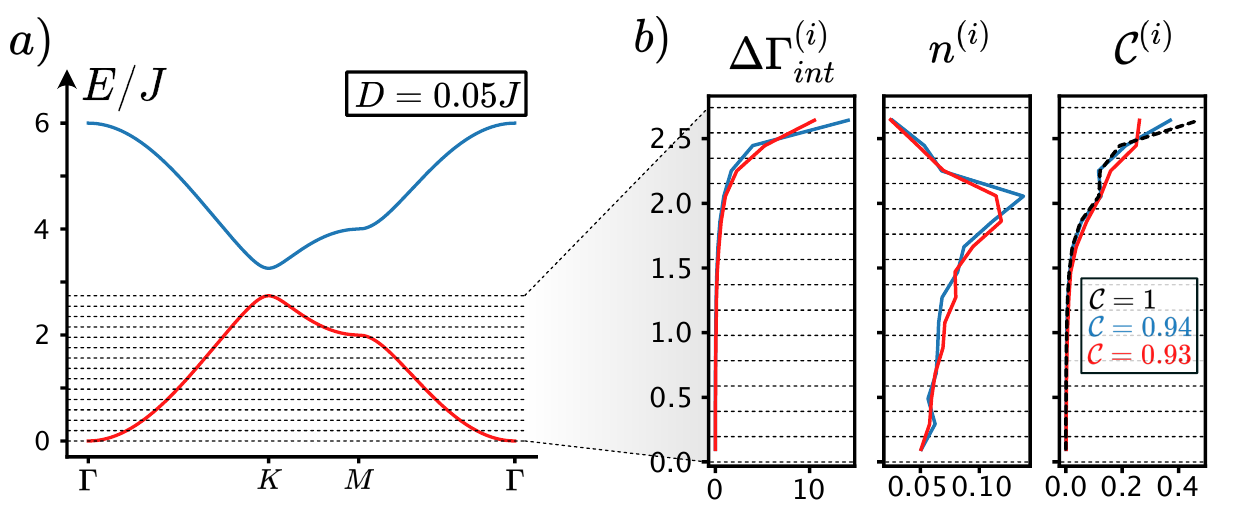}
    
    \caption{\textit{An energy resolved Chern marker measurement.} (a) Energy spectrum of the $2$D spin model in Eq.~\eqref{eq:spin_ham} for $D/J=0.05$ (b) The Lowest band bandwidth is separated into N=14 layers. The contribution of each layer $(i)$ to the local Chern marker $\mathcal{C}^{(i)}$ is given by the product of the integrated differential excitation rates for a state projected onto this layer $\Delta\Gamma^{(i)}_{int}$ by the density of states of the corresponding layer $n^{i}$. The blue curve represents the results obtained from the ideal preparation (projection of the initial state onto the energy layer), while the red curve represents the results obtained from the pump-dissipative protocol. In the rightmost figure, these measurements are compared to the theoretical results represented in a black dashed line. }
    \label{fig:layer}
\end{figure}
\textit{Energy-resolved measurement for dispersive bands ---} We now introduce a variant of the protocol by defining an ``energy-resolved Chern number", enabling a more accurate measurement of the Chern marker in systems with smaller bandgaps. Specifically, we partition the lowest band into $N$ energy layers and consider an initial single-magnon state that uniformly occupies one such layer while remaining localized on a single lattice site,
\begin{equation}
\label{eq:initial_spin_layer}
    \ket{\downarrow_i^{(j)}}=\frac{1}{\sqrt{n^{(j)}}}\!\!\sum_{g\in LB^{(j)}}\!\!\!\ket{g}\!\braket{g|i}\,,\,\textrm{with } n^{(j)}=\!\!\sum_{g\in LB^{(j)}} \!\!\left|\braket{g|i}\right|^2\,,
\end{equation}
where $LB^{(j)}$ defines the states of the lowest band whose energy belongs to the $j^{th}$ energy layer, and $n^{(j)}=n^{(j)} (\Vec{r}_i)$ is the local sub-band density of states. Extending the protocol developed in the previous paragraphs to each layer, we can then reconstruct the total Chern number layer by layer as

\begin{equation}
    \label{eq:reconstruction_layer}
    \frac{\hbar^2}{V_{\textrm{Cell}}\epsilon^2}\frac{\sum_jn^{(j)}(\Vec{r}_i)\Delta\Gamma^{(j)}(\Vec{r}_i)}{\sum_jn^{(j)}(\Vec{r}_i)}=\widetilde{\mathcal{C}}(\Vec{r}_i)\,,
\end{equation}
with $\Delta\Gamma^{(j)}(\Vec{r}_i)$ the integrated differential excitation rates associated to an initial state prepared at position $\Vec{r}_i$ in the $j^{\textrm{th}}$ layer. The layered measurement can then be interpreted by noting that $(n^{(j)}/\sum_kn^{(k)})\Delta\Gamma^{(j)}$ probes the Berry curvature of bands associated with layer $j$, thereby identifying Berry-curvature hotspots in the spectrum.

Within a pump–dissipative scheme, the layer-resolved initial state $|\!\downarrow_i^{(j)}\rangle$ can be prepared while simultaneously probing $n^{(j)}$. Specifically, one pumps at a frequency $\omega_0$ centered on the selected energy layer with a loss rate $\gamma$ set to one quarter of its width. For sufficiently narrow layers, the steady-state magnon number satisfies $N_{{\rm m}}\approx \frac{(\mu_BB_i)^2}{2\gamma^2S}n^{(j)}(\Vec{r}_i)$, so that measuring the pump power required to create a single magnon directly yields the sub-band density of states $n^{(j)}(\Vec{r}_i)$.

Simulations of this energy-resolved protocol for a Haldane model with $D/J=0.05$ (or $\Delta/W=0.19$),  corresponding to a gap that is roughly half the one observed in $CrI_3$,  were performed using 14 layers to decompose the lowest band (Fig.~\ref{fig:layer}). These results show good agreement between the exact value ($\mathcal{C}\!=\!1$), the estimate obtained from ideal initial states  ($\widetilde{\mathcal{C}}\!\approx\!0.94$), and the full pump–dissipative scheme ($\widetilde{\mathcal{C}}\!\approx\!0.93$). This substantial improvement can be understood by noting that the effective bandgap-to-bandwidth ratio is increased by a factor $N$ (i.e.~$\Delta/W_{{\rm eff}}=N\Delta/W$), thereby suppressing spurious excitations into the upper band. This protocol thus yields a more accurate estimate of the Chern marker while providing additional insight into the energy-resolved structure of the Berry curvature.\\


\textit{Discussion \& Perspectives ---} This work introduced a strategy to access a local topological marker in bosonic systems, which is based on combining a local driven-dissipative preparation scheme with a circular-dichroic measurement. While this work illustrated this scheme in the context of topological magnons, the recipe described above can be generalized to any spin system with bosonic low-energy excitations. In general, the protocol consists in identifying:~(i) a localized single-excitation state; (ii) a suitable driven-dissipative protocol to prepare this local state, while optimizing the uniform filling of the target topological band; and (iii) the chiral perturbation [Eq.~\eqref{eq:chiral_pert}], where the position operator takes the form $\hat{x}\!=\!\sum_ix_i\hat{n}_i$, with $\hat{n}_i$ the local bosonic densities.

An interesting perspective concerns the physics beyond the single-excitation approximation, which is considered in this work. It would be intriguing to explore the effectiveness of the preparation protocol beyond the classical mean-field theory, but also the effects of magnon-magnon interactions, and the fate of bulk topology in the absence of magnon conservation~\cite{habel2024breakdown}. Applying our method to light-induced topological magnons~\cite{Gregor_magnons} also emerges as a promising direction.

Systems in other symmetry classes and with different magnetic orders can also host Chern bands. A natural extension of this work is therefore to generalize our scheme to such systems—including noncollinear ferromagnets~\cite{fishman2025topological,dong2026topological}, antiferromagnets~\cite{owerre2017noncollinear,laurell2018magnon,yang2018quantum,mook2019thermal}, and altermagnets~\cite{khatua2025magnon}—by adapting the protocol to account for the geometry of para-unitary transformations arising from Bogoliubov–de Gennes Hamiltonians~\cite{Tesfaye}.

Beyond topological Chern bands, other types of topological properties can arise from spin systems, such as magnon (pseudo)-spin Hall states~\cite{kondo2019z,vinas2023direct} or Majorana fermions in Kitaev-type honeycomb models~\cite{kitaev2006anyons,sun2023engineering,koller2025raman}. In this context, our work paves the way to numerical and experimental investigations of real-space topological markers beyond Chern insulators, applicable to spin systems. The strategy described in this work can be generalized depending on the case under study:~by including a chiral modulation that depends on both the spin and the sublattice degree of freedom ($\hat{n}_i=\hat{\tau}^z_i\hat{\sigma}^z_i$) in the context of magnon spin-Hall systems, or by coupling the perturbation to the density of Majorana fermions in the case of the Kitaev honeycomb model.\\

\begin{paragraph}{Acknowledgments.}
 We thank Marco Schiro, Marin Bukov, Antoine Georges and Johannes Knolle for useful discussions.  This work was supported by the ERC (LATIS project), the EOS project CHEQS, and the Fondation ULB. We used Quspin for simulating the dynamics of the magnons and spin systems~\cite{10.21468/SciPostPhys.2.1.003,10.21468/SciPostPhys.7.2.020}
\end{paragraph}

\bibliographystyle{apsrev4-1}
\bibliography{Biblio}

\begin{thebibliography}{5}%
\makeatletter
\providecommand \@ifxundefined [1]{%
 \@ifx{#1\undefined}
}%
\providecommand \@ifnum [1]{%
 \ifnum #1\expandafter \@firstoftwo
 \else \expandafter \@secondoftwo
 \fi
}%
\providecommand \@ifx [1]{%
 \ifx #1\expandafter \@firstoftwo
 \else \expandafter \@secondoftwo
 \fi
}%
\providecommand \natexlab [1]{#1}%
\providecommand \enquote  [1]{``#1''}%
\providecommand \bibnamefont  [1]{#1}%
\providecommand \bibfnamefont [1]{#1}%
\providecommand \citenamefont [1]{#1}%
\providecommand \href@noop [0]{\@secondoftwo}%
\providecommand \href [0]{\begingroup \@sanitize@url \@href}%
\providecommand \@href[1]{\@@startlink{#1}\@@href}%
\providecommand \@@href[1]{\endgroup#1\@@endlink}%
\providecommand \@sanitize@url [0]{\catcode `\\12\catcode `\$12\catcode `\&12\catcode `\#12\catcode `\^12\catcode `\_12\catcode `\%12\relax}%
\providecommand \@@startlink[1]{}%
\providecommand \@@endlink[0]{}%
\providecommand \url  [0]{\begingroup\@sanitize@url \@url }%
\providecommand \@url [1]{\endgroup\@href {#1}{\urlprefix }}%
\providecommand \urlprefix  [0]{URL }%
\providecommand \Eprint [0]{\href }%
\providecommand \doibase [0]{http://dx.doi.org/}%
\providecommand \selectlanguage [0]{\@gobble}%
\providecommand \bibinfo  [0]{\@secondoftwo}%
\providecommand \bibfield  [0]{\@secondoftwo}%
\providecommand \translation [1]{[#1]}%
\providecommand \BibitemOpen [0]{}%
\providecommand \bibitemStop [0]{}%
\providecommand \bibitemNoStop [0]{.\EOS\space}%
\providecommand \EOS [0]{\spacefactor3000\relax}%
\providecommand \BibitemShut  [1]{\csname bibitem#1\endcsname}%
\let\auto@bib@innerbib\@empty
\bibitem [{\citenamefont {Bianco}\ and\ \citenamefont {Resta}(2011)}]{bianco2011mapping}%
  \BibitemOpen
  \bibfield  {author} {\bibinfo {author} {\bibfnamefont {R.}~\bibnamefont {Bianco}}\ and\ \bibinfo {author} {\bibfnamefont {R.}~\bibnamefont {Resta}},\ }\href@noop {} {\bibfield  {journal} {\bibinfo  {journal} {Physical Review B—Condensed Matter and Materials Physics}\ }\textbf {\bibinfo {volume} {84}},\ \bibinfo {pages} {241106} (\bibinfo {year} {2011})}\BibitemShut {NoStop}%
\bibitem [{\citenamefont {Breuer}\ and\ \citenamefont {Petruccione}(2002)}]{breuer2002theory}%
  \BibitemOpen
  \bibfield  {author} {\bibinfo {author} {\bibfnamefont {H.-P.}\ \bibnamefont {Breuer}}\ and\ \bibinfo {author} {\bibfnamefont {F.}~\bibnamefont {Petruccione}},\ }\href {\doibase https://doi.org/10.1093/acprof:oso/9780199213900.001.0001} {\emph {\bibinfo {title} {The theory of open quantum systems}}}\ (\bibinfo  {publisher} {OUP Oxford},\ \bibinfo {year} {2002})\BibitemShut {NoStop}%
\bibitem [{\citenamefont {Manzano}(2020)}]{manzano2020short}%
  \BibitemOpen
  \bibfield  {author} {\bibinfo {author} {\bibfnamefont {D.}~\bibnamefont {Manzano}},\ }\href {\doibase 10.1063/1.5115323} {\bibfield  {journal} {\bibinfo  {journal} {AIP Advances}\ }\textbf {\bibinfo {volume} {10}},\ \bibinfo {pages} {025106} (\bibinfo {year} {2020})}\BibitemShut {NoStop}%
\bibitem [{\citenamefont {Chisnell}\ \emph {et~al.}(2015)\citenamefont {Chisnell}, \citenamefont {Helton}, \citenamefont {Freedman}, \citenamefont {Singh}, \citenamefont {Bewley}, \citenamefont {Nocera},\ and\ \citenamefont {Lee}}]{chisnell2015topological}%
  \BibitemOpen
  \bibfield  {author} {\bibinfo {author} {\bibfnamefont {R.}~\bibnamefont {Chisnell}}, \bibinfo {author} {\bibfnamefont {J.~S.}\ \bibnamefont {Helton}}, \bibinfo {author} {\bibfnamefont {D.~E.}\ \bibnamefont {Freedman}}, \bibinfo {author} {\bibfnamefont {D.~K.}\ \bibnamefont {Singh}}, \bibinfo {author} {\bibfnamefont {R.~I.}\ \bibnamefont {Bewley}}, \bibinfo {author} {\bibfnamefont {D.~G.}\ \bibnamefont {Nocera}}, \ and\ \bibinfo {author} {\bibfnamefont {Y.~S.}\ \bibnamefont {Lee}},\ }\href {\doibase 10.1103/PhysRevLett.115.147201} {\bibfield  {journal} {\bibinfo  {journal} {Phys. Rev. Lett.}\ }\textbf {\bibinfo {volume} {115}},\ \bibinfo {pages} {147201} (\bibinfo {year} {2015})}\BibitemShut {NoStop}%
\bibitem [{\citenamefont {Hirschberger}\ \emph {et~al.}(2015)\citenamefont {Hirschberger}, \citenamefont {Chisnell}, \citenamefont {Lee},\ and\ \citenamefont {Ong}}]{hirschberger2015thermal}%
  \BibitemOpen
  \bibfield  {author} {\bibinfo {author} {\bibfnamefont {M.}~\bibnamefont {Hirschberger}}, \bibinfo {author} {\bibfnamefont {R.}~\bibnamefont {Chisnell}}, \bibinfo {author} {\bibfnamefont {Y.~S.}\ \bibnamefont {Lee}}, \ and\ \bibinfo {author} {\bibfnamefont {N.~P.}\ \bibnamefont {Ong}},\ }\href {\doibase 10.1103/PhysRevLett.115.106603} {\bibfield  {journal} {\bibinfo  {journal} {Phys. Rev. Lett.}\ }\textbf {\bibinfo {volume} {115}},\ \bibinfo {pages} {106603} (\bibinfo {year} {2015})}\BibitemShut {NoStop}%
\end{thebibliography}%
\begin{center}

{\bf End Matter}
\end{center}
\textit{Holstein-Primakoff transform and spin wave Hamiltonian ---}
A standard way to describe magnonic excitations above an ordered ground state is to map the spin operators to bosonic operators using a so-called Holstein–Primakoff
transformation~\cite{holstein1940field}. It is known that the ground state of our system is a ferromagnetic ground state with a global SO(3) symmetry~\cite{habel2024breakdown,dong2026topological}, provided that $D$ is small enough ($D\lesssim0.7JS$). The direction can then be fixed along the z-axis, without altering the total magnetization conservation, by adding a small magnetic field along the z-axis.
For such a polarized ground state, the Holstein-Primakoff transform takes the form 
\begin{align}
        &\hat{s}^+_i=\sqrt{2S}\sqrt{1-\frac{\hat{a}^\dagger_i\hat{a}_i}{2S}}\hat{a}_i , \quad 
        \hat{s}^-_i=\sqrt{2S}\hat{a}^\dagger_i\sqrt{1-\frac{\hat{a}^\dagger_i\hat{a}_i}{2S}},\label{eq:Holstein}
\end{align}
and $\hat{s}^z_i=S-\hat{a}^\dagger_i\hat{a}_i$, with $\hat{s}^\pm_i=\hat{s}^x_i\pm i\hat{s}^y_i$ and where $\hat{a}^\dagger_i$($\hat{a}_i$) creates (annihilates) a magnon at site $i$.

Plugging this transform in the Hamiltonian~\eqref {eq:spin_ham}, and performing a large-S expansion, the Hamiltonian becomes a power series in $1/S$ of the form
\begin{equation}
    \label{eq:large_s_H}
\hat{\mathcal{H}}=E_{GS}+\hat{\mathcal{H}}_2+\hat{\mathcal{H}}_4+\hat{\mathcal{H}}_6+...
\end{equation}
with $E_{GS}$ the classical ground state energy, and where $\hat{\mathcal{H}}_n$ contains $n$ bosonic operators $\hat{a}_i/\hat{a}^\dagger_i$ and is of order $S^{2-n/2}$.  
Note that the creation/annihilation operators only appear in equal numbers in each term of the Hamiltonian since the magnetization ($S_z$) is conserved. 
At the linear spin-wave level, the Hamiltonian~\eqref{eq:large_s_H} reads 
\begin{equation}
    \begin{aligned}
        \hat{\mathcal{H}}
        =&-N_lJS^2-JS\sum_{\langle ij\rangle}\left(\hat{a}^\dagger_j\hat{a}_i+\hat{a}^\dagger_i\hat{a}_j\right)\\&+JS\sum_iz_i\hat{a}^\dagger_i\hat{a}_i+iDS\sum_{\langle\langle ij\rangle\rangle}\eta_{ij}\left(\hat{a}^\dagger_j\hat{a}_i-\hat{a}^\dagger_i\hat{a}_j\right),
    \end{aligned}
\end{equation}
where $a_i^\dagger$($a_i$) creates (anihilates) a magnon at site i, with $N_l=\sum_iz_i$ the total number of links, and $z_i$ the number of nearest neighbors of site $i$, i.e., $z_i = 3$ in the bulk, and $z_i\le 3$ on the edges. \\

Building on this spin-boson mapping, we now express the bosonic position operator
\begin{equation}
    \hat{x}=\sum_ix_i\hat{a}^\dagger_i\hat{a}_i \, ,
\end{equation}
with $x_i$ the $x$-coordinate of site $i$, in terms of spin operators. To this end, we identify two possible solutions. 

A first possibility, as presented in the main text, consists in making use of the Holstein-Primakoff transform of $\hat{s}^z$ to rewrite the position operator as 
    \begin{equation}
    \label{app:spin_position_sz}
    \hat{x}=\sum_ix_i(S-\hat{s}^z_i)\,.
\end{equation}
Since all the matrix elements involved in the Fermi's golden rules are of the form $\braket{e|\hat{x}|g}$, with $\ket{e}$ and $\ket{g}$ two orthogonal states, the expression of the position
operator~\eqref{app:spin_position_sz} can further be 
simplified to: $\hat{x}=-\sum_ix_i\hat{s}^z_i$; see Eq.~\eqref{x_spin}.

We note that this simplification is only valid in the ideal limit where only the lowest-band states are excited at $t=0$; beyond this limit, the full expression~\eqref{app:spin_position_sz} is required.

Another approach would be to use the mapping ``$\hat{a}_i^\dagger\to\hat{s}_i^-$'' and ``$\hat{a}_i\to\hat{s}_i^+$'', such that we write the spin position operator as ~\cite{ozawa2019probing}
\begin{equation}
\label{eq:mapping_pos_+-}
    \hat{x}=\frac{1}{2S}\sum_jx_j\hat{s}_j^-\hat{s}_j^+\,.
\end{equation}
Based on the spin algebra $\hat{s}^-\hat{s}^+=S(S+1)\hat{1\!\!1}-(\hat{s}^z)^2-\hat{s}^z$, we identify two cases where the replacement $\hat{a}^\dagger \hat{a}\to \hat{s}^-\hat{s}^+$ is valid: first, for spin-1/2 particles, the identity $(\hat{s}^z)^2=S^2\hat{1\!\!1}$ leads to $\hat{s}^-\hat{s}^+=\frac12-\hat{s}^z$; and second, the large S-limit, in which Eq.~\eqref{eq:Holstein} implies $\frac{1}{2S}\hat{s}^-\hat{s}^+=\hat{a}^\dagger \hat{a}-\frac{\hat{a}^\dagger \hat{a}^\dagger \hat{a} \hat{a}}{2S}\approx \hat{a}^\dagger \hat{a}$. \\ 

\textit{Fermi's Golden rules and local Chern markers ---}
Assuming a chiral drive [Eq.~\eqref{eq:chiral_pert}] of small amplitude, we evaluate the excited fraction in the higher band at first order in time-dependent perturbation theory, and obtain
\begin{equation}
\label{eq:density}
\scalemath{0.83}{
    N^\pm_{i}(\omega,t) \smeq \frac{\epsilon^2}{\hbar^2}\sum_{e\notin LB}\left|\sum_{g\in LB}\bra{e}\hat{x}\pm i\hat{y}\ket{g} \braket{g|\downarrow_i}\frac{e^{it(\omega_{eg}-\omega)}-1}{\omega_{eg}-\omega}\right|^2} ,
\end{equation}
where $\hbar\omega_{eg}=E_e-E_g$ , and $\ket{e}$ denotes states belonging to the higher-energy band [Fig.~\ref{fig:bosonized_results}(a)].\\
The integrated circular dichroic signal then results in
\begin{align}
\Delta\Gamma_i(t)&=\int_0^{+\infty}\textrm{d}\omega\frac{N_i^{+}(\omega,t)-N_i^{-}(\omega,t)}{2t} \\
&\underset{t\to\infty}{=}-\frac{4\pi \epsilon^2}{\hbar^2}{\rm Im}\left(\sum_{g\in LB}\braket{ \downarrow_i|g}\langle g|\hat{x}\hat{Q}\hat{y}|g\rangle\braket{ g|\downarrow_i}\right) \,, \notag 
\end{align}
from which we define a Chern marker 
\begin{align}
    \widetilde{\mathcal{C}}(\Vec{r}_i)&=\underset{t\to \infty}{\textrm{lim}}\left(\frac{\hbar^2}{V_{{\rm cell}}\epsilon^2}\Delta\Gamma_i\right)\\
    &=-\frac{4\pi}{V_{{\rm cell}}}{\rm Im}\left(\sum_{g\in LB}\braket{ \downarrow_i|g}\bra{g}\hat{x}\hat{Q}\hat{y}\ket{g}\braket{g|\downarrow_i}\right)\,. \notag 
\end{align}
For comparison, we recall the expression of the Chern marker introduced by Bianco and Resta~\cite{bianco2011mapping}:
\begin{align}
\label{eq:mark_bianco_resta}
    \mathcal{C}(\Vec{r}_i)&=-\frac{4\pi}{V_{{\rm cell}}}{\rm Im}\left(\bra{ \downarrow_i}\hat{P}\hat{x}\hat{Q}\hat{y}\hat{P}\ket{\downarrow_i}\right) \\
    &=-\frac{4\pi}{V_{{\rm cell}}}{\rm Im}\sum_{g,g'\in LB}\braket{ \downarrow_i|g}\bra{g}\hat{x}\hat{Q}\hat{y}\ket{g'}\braket{g'|\downarrow_i}\,. \notag 
\end{align}
While the two markers differ through the off-diagonal terms of the form $\braket{ g|\hat{x}\hat{Q}\hat{y}|g'}$, with $g$ and $g'$ two different states belonging to the lowest band, their values are similar as illustrated in Fig.\ref{fig:bosonized_results}(d). This resemblance arises because, in the bulk, the system approximates one with periodic boundary conditions, where quasi-momenta $\Vec{k}$ are good quantum numbers and the lowest band states are momentum eigenstates. Since $\hat{x}\equiv\frac{i}{\hbar}\partial_{kx}$ is diagonal in momentum space, so is $\hat{x}\hat{Q}\hat{y}$, implying $\braket{ g|\hat{x}\hat{Q}\hat{y}|g'}=\delta_{g,g'}\braket{ g|\hat{x}\hat{Q}\hat{y}|g}$. After summation over $g'$ one therefore recovers $\widetilde{\mathcal{C}}(\Vec{r}_i)=\mathcal{C}(\Vec{r}_i)$.\\
 
\textit{Numerical approximations ---}
The Chern marker measurement provided in this manuscript relies on two key assumptions (discussed in detail in~\cite{SuppMat}). First, the relationship between the marker and differential excitation rates involves a frequency integral, which, in any numerical or experimental measurement, must be approximated by a discrete sum over a finite set of frequencies~\cite{asteria2019measuring}, with spacing $\Delta\omega$, such that the integral in Eq.~\eqref{eq:rate_diff} is approximated by:
\begin{equation}
\Delta\omega\sum_{n\geq0} \left[ \Gamma_{i}^+(n\Delta\omega,t)-\Gamma_{i}^-(n\Delta\omega,t) \right] \, .
\end{equation}
Such a discretization remains valid as long as $t\!<\!2\pi/\Delta\omega$, although the achievable precision also depends on the number of frequency points.

Second, the perturbative expression in Eq.~\eqref{eq:density} holds only for small excitation amplitudes $\epsilon$, ensuring $N_{i}^{\pm}(\omega,t)\ll1$ for all relevant $\omega$ and $t$. Since the relationship between chiral excitation rates and the local Chern marker Eq.~\eqref{eq:Chern_Mark_us} is derived within this regime, a proper choice of $\epsilon$ is essential.\\

\clearpage
\setcounter{equation}{0}
\setcounter{figure}{0}
\setcounter{table}{0}
\makeatletter
\renewcommand{\theequation}{S\arabic{equation}}
\renewcommand{\thefigure}{S\arabic{figure}}
\renewcommand{\thetable}{S\arabic{table}}
\renewcommand{\theHequation}{S\arabic{equation}}
\renewcommand{\theHfigure}{S\arabic{figure}}
\renewcommand{\theHtable}{S\arabic{table}}
\makeatother
\clearpage
\setcounter{equation}{0}
\setcounter{figure}{0}
\setcounter{table}{0}

\onecolumngrid
\setcounter{page}{1}
\makeatletter
\renewcommand{\theequation}{S\arabic{equation}}
\renewcommand{\thefigure}{S\arabic{figure}}
\renewcommand{\bibnumfmt}[1]{[S#1]}
\renewcommand{\citenumfont}[1]{S#1}
\begin{center}
\textbf{\large SUPPLEMENTAL MATERIAL\\
A local quantized marker for topological magnons from circular dichroism}
\end{center}
\begin{bibunit}  
    \section{Time-dependent perturbation theory and numerical limitations }

In this section of the supplemental materials, we revisit time-dependent perturbation theory and circular dichroism to derive Eq.~\eqref{eq:rate_diff} employed in the main text and Eq.~\eqref{eq:density} of the End Matter.\\

\subsection{Time-dependent perturbation theory}
\label{app:time_dep_pert}
Let us consider a Hamiltonian $\hat{\mathcal{H}}_0$ of eigenstates $|n\rangle$, such that $\hat{\mathcal{H}}_0|n\rangle=E_n|n\rangle=\hbar\omega_n|n\rangle$, and a time-dependent perturbation of the form \begin{equation}
    \delta\hat{\mathcal{H}}_\pm(t)=2\epsilon\left[\hat{x}\cos\left(\omega t\right)\pm\hat{y}\sin\left(\omega t\right)\right]\,.
\end{equation}
To proceed, we express the system's wavefunction in the interaction picture in the form
\begin{equation}
    \left|\Psi(t)\right\rangle=\sum_nc_n(t)e^{-i\omega_nt}\left|n\right\rangle\,,
\end{equation}
such that the Schr\"odinger equation 
\begin{equation}
    i\hbar\partial_t\left|\Psi(t)\right\rangle=\left(\hat{\mathcal{H}}_0+\delta\hat{\mathcal{H}}_\pm(t)\right)\left|\Psi(t)\right\rangle
\end{equation}
projected onto the eigenstates of $\hat{\mathcal{H}}_0$ takes the form
\begin{equation}
    i\hbar\partial_tc_n(t)=\sum_mc_m(t)e^{-i(\omega_m-\omega_n)t}\left\langle n\left|\delta\hat{\mathcal{H}}_\pm(t)\right|m\right\rangle \, , \quad  \forall \, n \, .
\end{equation}
At linear order in $\epsilon$ in perturbation theory, the coefficients $c_n(t)$ are given by
\begin{equation}
    c_n(t)=c_n(0)+\epsilon\sum_mc_m(0)\left(\frac{e^{-i(\omega_m-\omega_n-\omega)t}-1}{E_m-E_n-\hbar\omega}\left\langle n\left|\hat{x}\mp i\hat{y}\right|m\right\rangle+\frac{e^{-i(\omega+\omega_m-\omega_n)t}-1}{E_m-E_n+\hbar\omega}\left\langle n\left|\hat{x}\pm i\hat{y}\right|m\right\rangle\right)\,.
    \label{eq_linear_epsilon}
\end{equation}
Assuming, as in the main text, that the system is initialized in a state $\ket{\psi_0}$ at $t=0$ such that only the lowest band is occupied (\textit{i.e.} $\forall n\notin LB\,,c_n(0)=\langle n|\psi_0\rangle=0$), the density of particles in the higher bands corresponds to the number of excited particles from the lowest bands to the others between times $0$ and $t$. It satisfies
\begin{equation}
        \begin{aligned}
            N^{\pm}(\omega,t)&=\sum_{n\notin LB}\left|c_n(t)\right|^2\\
            &=\epsilon^2\sum_{n\notin LB}\left|\sum_{m\in LB}c_m(0)\left(\left\langle n\left|\hat{x}\mp i\hat{y}\right|m\right\rangle \frac{e^{-it(\omega_m-\omega_n-\omega)}-1}{E_m-E_n-\hbar\omega}+\left\langle n\left|\hat{x}\pm i\hat{y}\right|m\right\rangle \frac{e^{-it(\omega_m-\omega_n+\omega)}-1}{E_m-E_n+\hbar\omega}\right)\right|^2\,.
        \end{aligned}
\end{equation}
Under a rotating wave approximation, valid in the regime $\omega t\gg2\pi$, and given that $\omega_m<\omega_n$, the number of particles in the excited states then reads
\begin{equation}
\label{eq:npm}
       N^{\pm}(\omega,t)=\left(\frac{\epsilon}{\hbar}\right)^2\sum_{n\notin LB}\left|\sum_{m\in LB}\left\langle n\left|\hat{x}\pm i\hat{y}\right|m\right\rangle\left\langle m|\psi_0\right\rangle \frac{e^{it(\omega_n-\omega_m-\omega)}-1}{\omega_n-\omega_m-\omega}\right|^2 \, ,
\end{equation}
which is identical to Eq.~\eqref{eq:density} in the End Matter.\\

\subsection{Circular dichroism and local Chern marker}

Because we consider an initial state localized in position $\ket{\psi_0}$, its occupations in the lowest band states $\langle m|\psi_0\rangle$, $\left\{ m \in LB \right\}$, are correlated, leading to interferences between eigenstates of different energies. 
One can get rid of these interferences by integrating the excitation rates $N_\pm(\omega,t)/t$ over all relevant frequencies. 
Using the limit,
\begin{equation}
\label{eq:dirac}
    \frac{1}{t}\int \frac{e^{it(\omega_1-\omega)}-1}{\omega_1-\omega}\frac{e^{-it(\omega_2-\omega)}-1}{\omega_2-\omega}d\omega\underset{t\to+\infty}{\longrightarrow}2\pi\delta_{\omega_1,\omega_2}\,,
\end{equation}
valid for $t\gg\frac{2\pi}{\omega_1-\omega_2}$, one gets
\begin{equation}
    \int d\omega\Gamma^{\pm}(\omega,t)\underset{t\to+\infty}{\longrightarrow}\frac{2\pi \epsilon^2}{\hbar^2}\sum_{e\notin LB}\sum_{g\in LB}\left|\left\langle e\left|\hat{x}\pm i\hat{y}\right|g\right\rangle\right|^2\left|\left\langle g|\psi_0\right\rangle\right|^2\,,
\end{equation}
such that the difference of the chiral excitation rates reads
\begin{equation}
\label{eq:Marker_Us}
    \begin{aligned}
        \int d\omega\left(\Gamma^{+}(\omega,t)-\Gamma^{-}(\omega,t)\right)&\underset{t\to+\infty}{\longrightarrow}-\frac{8\pi \epsilon^2}{\hbar^2}Im\left(\sum_{e\notin LB}\sum_{g\in LB}\left|\left\langle \psi_0|g\right\rangle\right|^2\left\langle g\left|\hat{x}\right|e\right\rangle\left\langle e\left|\hat{y}\right|g\right\rangle\right)\\
        &\underset{t\to+\infty}{\longrightarrow}-\frac{8\pi \epsilon^2}{\hbar^2}Im\left(\sum_{g\in LB}\left\langle \psi_0|g\right\rangle\langle g|\hat{x}\hat{Q}\hat{y}|g\rangle\left\langle g|\psi_0\right\rangle\right) \, ,
    \end{aligned}
\end{equation}
where $\hat{Q}$ is the projector onto the highest bands $\hat{Q}=\sum_{e\notin LB}\left|e\right\rangle\left\langle e\right|$, recovering Eq.~\eqref{eq:rate_diff} of the main text.\\

\subsection{Limitations of the time-dependent perturbation theory analysis}

We aim here to review the various assumptions necessary to justify the link between the local quantized Chern marker and the differential integrated excitation rate, as expressed in Eqs.~\eqref{eq:rate_diff} and \eqref{eq:Chern_Mark_us}, within the framework of linear response theory. \\ 

\begin{figure}
    \centering
    \includegraphics[width=0.5\linewidth]{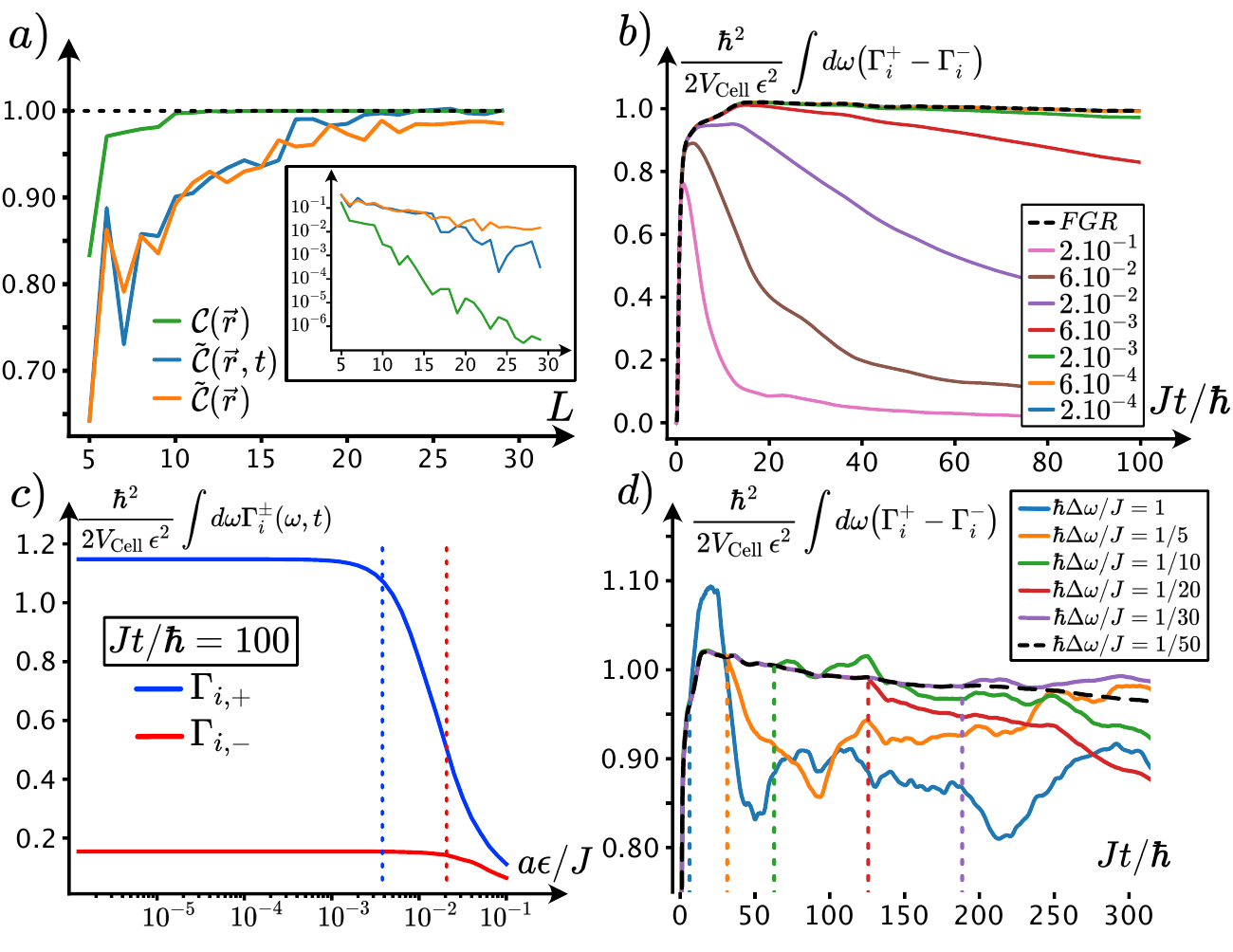}
    
    \caption{\textit{Limitations of the local Chern markers.} (a) Finite-size analysis of the three types of local Chern markers discussed in the article: we compare the standard local Chern marker introduced by Bianco and Resta in~\cite{bianco2011mapping} [Eq.~\eqref{eq:mark_bianco_resta}] (green), the finite time estimation of the local Chern marker introduced in this work $\widetilde{\mathcal{C}}(\Vec{r},t)$ at $Jt/\hbar=100$ (blue), and its long-time limit $\lim_{t\to\infty} \widetilde{\mathcal{C}}(\Vec{r},t)$ (orange) [Eq.~\eqref{eq:Chern_Mark_us}], as functions of the system size. The inset shows the same data on a semi-logarithmic scale to highlight the exponential convergence of the markers. (b) Time evolution of the integrated differential rate [Eq.~\eqref{eq:rate_diff}] for different perturbation amplitudes $\epsilon$. As $\epsilon$ decreases, the dynamics converge toward the value predicted by Fermi’s golden rule, represented in dashed lines. (c) The integrated chiral excitation rates $\int\text{d}\omega \, \Gamma^{\pm}(\omega,t)$ at $Jt/\hbar=100$ are shown as functions of the excitation amplitude $a \,\epsilon/J$. The dashed lines represent the critical excitation amplitudes $\epsilon_c$ [Eq.~\eqref{eq:critical_amplitude}]. (d) Time evolution of the integrated differential rate [Eq.~\eqref{eq:rate_diff}] displayed for several integration steps $\Delta \omega$. The curves with larger $\Delta \omega$ begin to deviate from the ideal integral ($\sim \Delta \omega\to0$) around $Jt/\hbar=2\pi/\Delta \omega$, as marked by the vertical dashed lines.}
    \label{fig:app_converge}
\end{figure}

First, we note that the first-order approximation in $\epsilon$ of the perturbative expansion for the excited fraction [Eqs.~\eqref{eq_linear_epsilon}-\eqref{eq:npm}] holds provided that, for any initial state in the lowest band, 
\begin{equation}
1\gg\frac{2\pi\epsilon^2 t}{\hbar^2}\left(|\braket{e|\hat{x}\pm i\hat{y}|g}|^2\right)\,.
\end{equation}
When this hypothesis is relaxed, the integrated differential excitation rate $\Delta\Gamma$ differs from Fermi golden rule's prediction, and it decreases in the long-time limit, as confirmed in Fig.~\ref{fig:app_converge}(b). Additionally, Fig.~\ref{fig:app_converge}(c) illustrates that both chiral excitation rates $\Gamma_{i}^{\pm}(t)$ reach a constant value when the time $t$ is fixed and $\epsilon$ remains below the critical value \begin{equation}
\label{eq:critical_amplitude}\epsilon_c^{-2}\approx\frac{2\pi t}{\hbar^2}\underset{g\in LB,\,e\notin LB}{\textrm{max}}\left(|\braket{e|\hat{x}\pm i\hat{y}|g}|^2\right)\,.
\end{equation}. \\ 

Moreover, the integral is approximated by a finite sum, with a frequency step $\Delta \omega$, as is done in any reasonable numerical or experimental situation. Such an approximation, however, only holds when $t$ satisfies $\Delta \omega t \ll 2\pi$. This constraint is exemplified in Fig.~\ref{fig:app_converge}(d), where we observe that the curves obtained for larger values of $\Delta\omega$ begin to deviate from the others around $2\pi/\Delta\omega$, indicated by the dashed line. One must therefore balance the three key timescales imposed by (i) the measurement duration, (ii) the finite perturbation strength $\epsilon$, and (iii) the energy resolution we wish to probe, as described in Eq.~\eqref{eq:dirac}. Achieving finer energy resolution requires longer measurement times, which in turn necessitates smaller excitation amplitudes and a greater number of measurements across different frequencies.

\newpage\section{Initial state preparation}

\begin{figure}
    \centering
    \includegraphics[width=0.75\linewidth]{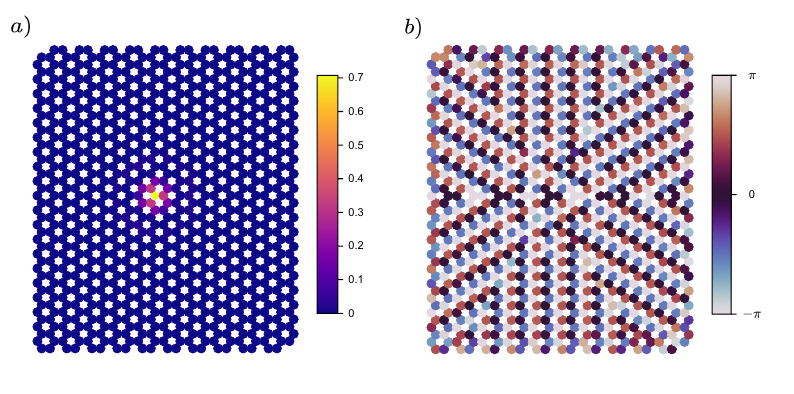}
    
    \caption{\textit{Representation of the theoretical 
    initial state} [Eq.~\eqref{eq:initial_spin}]. (a) Magnon density for an initial state centered in the middle of the sample and (b) the corresponding phases.}
\label{fig:initial_state}
\end{figure}

In this letter, we discuss a measurement protocol to access a local quantized marker for topological magnons based on the principle of circular dichroism. However, because of the bosonic nature of the excitations, the ground state of the system consists of an empty vacuum state. It is therefore crucial for the measurement to first excite a localized initial state, which entirely projects
onto a single topological band. \\

In the main text, we propose to consider as an initial state $\ket{\downarrow_i^s}$, the projection onto the lowest band of a single-spin-flip excitation described by Eq.~\eqref{eq:initial_spin}. 

While the spin-flip is strictly localized on a single lattice site, after the projection onto the lowest band, $\ket{\downarrow_i^s}$, the initial state presents features beyond the excited spin-site. In particular, defining $\ket{j}\equiv\hat{s}_j^-\ket{0}$, the initial state can be decomposed on all lattice sites as 
\begin{equation}
    \ket{\downarrow_i^s}=\sum_j\ket{j}\braket{j|\downarrow_i^s}\, ,
\end{equation}
where the norm and the phase associated to the overlap $\braket{j|\downarrow_i^s}$ are represented respectively in Fig.~\ref{fig:initial_state} a) and b).

\subsection{Lindblad master equation formalism for the pump-dissipative preparation scheme}
\label{sec:Lindblad}
In the main text, we discuss the possibility of realizing the targeted initial state using a pump-dissipation protocol. In this section, we will then review the theoretical tools leading to the mean-field equations in the presence of both losses and pumping used in the main text for the initial state preparation.\\

In order to describe the loss rate in the spin system, we resort to a Lindblad master equation for the reduced density matrix $\hat{\rho}$ \cite{breuer2002theory,manzano2020short}
\begin{equation}
    i\hbar\partial_t\hat{\rho}=\left[\hat{\mathcal{H}},\hat{\rho}\right]+i\hbar\sum_k 2\gamma_k\left(\hat{L}_k\hat{\rho}\hat{L}_k^\dagger-\frac{1}{2}\left\lbrace \hat{L}_k^\dagger \hat{L}_k,\hat{\rho}\right\rbrace\right) \, ,
\end{equation}
with $\hat{\mathcal{H}}$ the total Hamiltonian of the system with pumping at a frequency $\omega_0$ 
\begin{equation}
    \hat{\mathcal{H}}=-J\sum_{\langle ij\rangle}\Vec{s}_i\cdot\Vec{s}_j+\sum_{\langle\langle ij\rangle\rangle}\Vec{D}_{ij}\cdot\left(\Vec{s}_i\wedge\Vec{s}_j\right)+\sum_i\mu_BB_i\left(e^{i\omega_0t}\hat{s}^+_i+e^{-i\omega_0t}\hat{s}^-_i\right)\,,
\end{equation}
and where $\hat{L}_k$ denotes a set of jump operators in our case given by $\hat{L}_k=\hat{s}^+_k$. Assuming, as in the main text, homogeneous losses in the overall system  ($\gamma_k\equiv\gamma$), each operator $\hat{\mathcal{O}}$ in the Heisenberg picture evolves as 
\begin{equation}
    i\hbar\partial_t\hat{\mathcal{O}}=\left[\hat{\mathcal{H}},\hat{\mathcal{O}}\right]+i\hbar\sum_k 2\gamma_k\left(\hat{L}_k\hat{\mathcal{O}}\hat{L}_k^\dagger-\frac{1}{2}\left\lbrace \hat{L}_k^\dagger \hat{L}_k,\hat{\mathcal{O}}\right\rbrace\right)\,.
\end{equation}
As a consequence, the Heisenberg equations for the spin operators $\hat{s}^z_i/\hat{s}^\pm_i$ read
\begin{subequations}
   \begin{align}
       i\hbar\partial_t\hat{s}_i^z&=\left[\hat{\mathcal{H}},\hat{s}^z_i\right]+2i\hbar\gamma\hat{s}_i^-\hat{s}_i^+\,,\\
       i\hbar\partial_t\hat{s}_i^\pm&=\left[\hat{\mathcal{H}},\hat{s}^\pm_i\right]-2i\hbar\gamma\hat{s}_i^z\hat{s}_i^\pm\,.
   \end{align} 
\end{subequations}
leading to the mean-field equations
\begin{subequations}
\label{eq:mean_field_spin}
    \begin{align}
       i\partial_ts^z_i=&-\frac{J}{2}\sum_{\langle ij\rangle}\left(s^+_is^-_j-s^-_is^+_j\right)+\frac{iD}{2}\sum_{\langle\langle ij\rangle\rangle}\eta_{ij}\left(s^+_is^-_j+s^-_is^+_j\right)+\mu_BB_i\left(e^{ i\omega_0t}s^+_i-e^{- i\omega_0t}s^-_i\right)+2i\gamma s_i^-s_i^+\,,\\
       \pm i\partial_ts^\pm_i=&-J\sum_{\langle ij\rangle}\left(-s^\pm_is^z_j+s^z_is^\pm_j\right) \mp iD\sum_{\langle\langle ij\rangle\rangle}\eta_{ij}s^z_is^\pm_j+2\mu_BB_ie^{\mp i\omega_0t}s^z_i\mp 2i\gamma s^z_is^\pm_i\,,
    \end{align}
\end{subequations}
where $s_i^z=\braket{\hat{s}^z_i}$ and $s_i^\pm=\braket{\hat{s}^\pm}$.\\
Working in the rotating wave picture around the z-axis, setting  $\sigma^\pm_i=e^{\pm i\omega_0t}s^\pm_i$, mean-field equations~\eqref{eq:mean_field_spin} can be expressed as
\begin{subequations}
\label{eq:mean_field_spin_rotating}
    \begin{align}
       i\partial_ts^z_i=&-\frac{J}{2}\sum_{\langle ij\rangle}\left(\sigma^+_i\sigma^-_j-\sigma^-_i\sigma^+_j\right)+\frac{iD}{2}\sum_{\langle\langle ij\rangle\rangle}\eta_{ij}\left(\sigma^+_i\sigma^-_j+\sigma^-_i\sigma^+_j\right)+\mu_BB_i\left(\sigma^+_i-\sigma_i^-\right)+2i\gamma\sigma_i^-\sigma_i^+\,,\\
       \pm i\partial_t\sigma^\pm_i=&-J\sum_{\langle ij\rangle}\left(-\sigma^\pm_is^z_j+s^z_i\sigma^\pm_j\right) \mp iD\sum_{\langle\langle ij\rangle\rangle}\eta_{ij}s^z_i\sigma^\pm_j+2\mu_BB_is^z_i-\left(\omega_0\pm 2i\gamma s^z_i\right)\sigma^\pm_i\,.
    \end{align}
\end{subequations}
Eq.~\eqref{eq:mean_field_large}, can be deduced directly by applying the Holstein-Primakoff transform [Eq.~\eqref{eq:Holstein}] to the spin operators at the linear order in the creation/annihilation operators.\\

In the main text, we indicate the mean magnon number for a uniformly populated layer
\begin{equation}
    \label{eq:density_magnon}N_m\approx\frac{(\mu_BB_i)^2}{2\gamma^2S}n ,
\end{equation}
with $n$ the layer density of states. This result can be obtained by building on Eq.~\eqref{eq:mean_field_large}. Indeed, considering the mean field equation for an eigenstate $\alpha$ of the system, of eigenenergy $\varepsilon_\alpha$, its occupation $n_\alpha$ verifies
\begin{equation}
    n_\alpha=\frac{2S\mu_B^2B_i^2}{(\varepsilon_\alpha-\omega_0)^2+4\gamma^2 S^2} .
\end{equation}
Such that, assuming a homogeneous filling of the layer, of energy close to $\epsilon_\alpha-\omega_0$, 
\begin{equation}
    n_\alpha\approx\frac{\mu_B^2B_i^2}{2\gamma^2S} ,
\end{equation}
from which results Eq.~\eqref{eq:density_magnon}.

\subsection{Beyond the single spin pump}
\begin{figure}[h]
    \centering
    \includegraphics[width=0.7\linewidth]{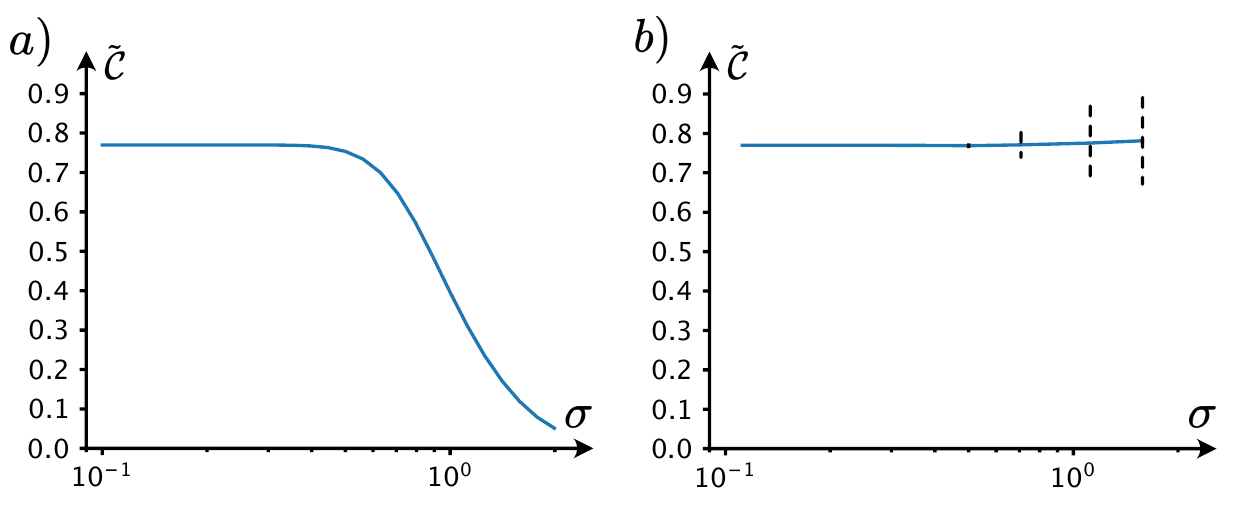}
    \caption{\textit{Local Chern marker obtained with a Gaussian pump profile.} (a) Local Chern marker obtained with a Gaussian pump $B_i(r_j)\propto e^{-|r_i-r_j|^2/2\sigma^2}$ as a function of the Gaussian width $\sigma$. (b) Average local Chern marker obtained with an incoherent Gaussian pump, as a function of the Gaussian width $\sigma$. The dashed line represents an estimation of the standard deviation of the measurement obtained for 20 different realizations of the preparation protocol.}
    \label{fig:gaussian}
\end{figure}

In the main text, the state preparation is introduced through a minimal model, where the magnetic field generates spin excitations on a single site $i$: $B_j \propto \delta_{ij}$. While state-of-the-art experimental techniques can realize single-site spin excitations (see main text), we nonetheless investigate here the fate of the quantized Chern marker in the presence of a spatially extended pump. \\ 

We apply the same measurement protocol as described in the main text, but now considering a Gaussian magnetic profile centered deep in the bulk:
\begin{equation}
    B_j\propto e^{-\frac{|\Vec{r}_i-\Vec{r}_j|^2}{2\sigma^2}}\,.
\end{equation}
As shown in Fig.~\ref{fig:gaussian}(a), the Chern marker remains well-quantized for small magnetic spot sizes ($\sigma\lesssim0.5$), where only a single site is significantly populated, recovering the results of the ideal (Dirac) pump protocol. For larger $\sigma$, the Gaussian profile becomes increasingly narrow in momentum space, causing the measurement to probe the vanishing Berry curvature near the $\Gamma$-point, which progressively destroys the quantization.

From the understanding of this effect, it is then possible to devise a strategy allowing one to preserve the quantization of the Chern marker while increasing the size of the magnetic field spot $\sigma$. This strategy consists of employing an incoherent magnetic field. Numerically, such a protocol is realized by adding to the pumping field a uniformly distributed random phase $\theta_j \in [0,2\pi[$ across the lattice, such that
\begin{equation}
    B_j\propto e^{-\frac{|\Vec{r}_i-\Vec{r}_j|^2}{2\sigma^2}+i\theta_j}\,.
\end{equation}
Averaging over such an incoherent field then destroys the phase coherence between neighboring spins, thus amounting to effectively compute the weighted average
\begin{equation}
  \Bar{\mathcal{C}}_i=\frac{\sum_j e^{-\frac{|\Vec{r}_i-\Vec{r}_j|^2}{\sigma^2}}\tilde{\mathcal{C}}_j}{\sum_j e^{-\frac{|\Vec{r}_i-\Vec{r}_j|^2}{\sigma^2}}}\,,  
\end{equation}
which in the bulk coincides with $\tilde{\mathcal{C}}_i$. This is illustrated in Fig.~\ref{fig:gaussian}(b) for an average over 20 phase realizations. We note, however, that the variance increases with the spot size (dashed lines). This indicates that probing such systems would require a larger number of measurements.
\section{Beyond two bands: Circular dichroism in a Kagome ferromagnet}
\label{sec:3bands}
\begin{figure}[h]
    \centering
    \includegraphics[width=0.9\linewidth]{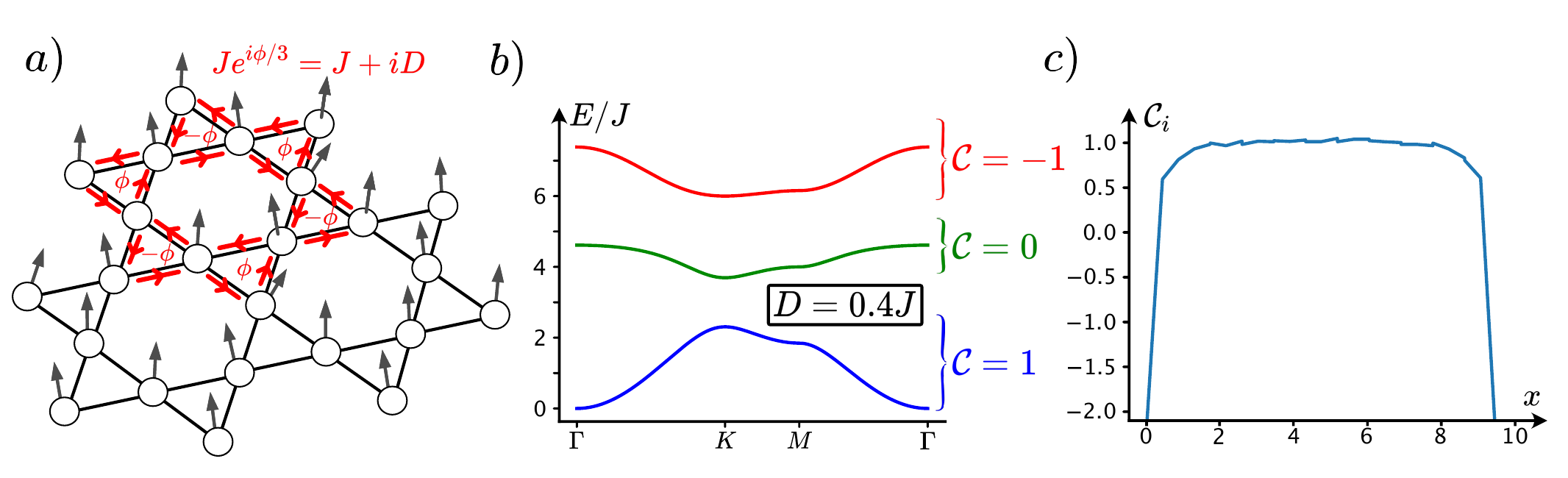}
    \caption{\textit{Local Chern marker measurement on the Kagome Haldane model.} (a) Illustration of the $2$D spin model [Eq.~\eqref{eq:kagome_spin}] with nearest-neighbor ferromagnetic Heisenberg and DM interactions. (b) The corresponding energy spectrum displays three bands, with Chern numbers $\mathcal{C}=0,\pm1$. Here we set $D=0.4J>0$. (c) The local Chern marker $\tilde{\mathcal{C}}_i$ exhibits a quantized value to $1$ in the bulk, while it takes large negative values near the edges.}
    \label{fig:3bands}
\end{figure}

While the main text focuses on a minimal two-band setting, the dichroic protocol naturally generalizes to systems with more bands. We illustrate this here on a 3-band topological magnon model.
\subsection{Kagome Haldane model}
Let us consider another prototypical model of topological magnon system that captures the low-energy magnon band of \iupac{Cu[1,3-benzene|di|carboxylate]}~\cite{chisnell2015topological,hirschberger2015thermal}. This model consists of a ferromagnetic material with spin-orbit coupling, whose localized spins are located on a Kagome lattice. The Hamiltonian is expressed as [Fig.~\ref{fig:3bands}(a)]:
\begin{equation}
\label{eq:kagome_spin}
    \hat{\mathcal{H}}=\sum_{\braket{ij}} \left[ -J\vec{s}_i\cdot\vec{s}_j+\vec{D}_{ij}\left(\vec{s}_i\wedge\vec{s}_j\right) \right] \,,
\end{equation}
where the spin operators $\vec{s}_i$ describe spin-degrees of freedom. The first term of \eqref{eq:kagome_spin} is a nearest-neighbor Heisenberg interaction, and we set $J>0$ to stabilize the ferromagnetic order. The second term is the nearest-neighbor DM interaction, which, based on the symmetries of the model, does not have any in-plane component ($\Vec{D}_{ij}=D_{ij}\Vec{e}_z$). Considering a uniform system, we set $D_{ij}=-D_{ji}=D\eta_{ij}$ where the sign $\eta_{ij}$ depends on the orientation of the two nearest-neighbor spins, as shown in red in Fig.~\ref{fig:3bands}(a).\\
To describe the magnonic excitations above the ferromagnetic ordered ground states (polarized along the $z-$axis by adding a small magnetic field), we employ a Holstein-Primakoff transform. Restraining ourselves to linear spin waves, one obtains 
\begin{equation}
    \label{eq:kagome}
    \hat{\mathcal{H}}=-N_lJS^2-JS\sum_{\langle ij\rangle}\left(\hat{a}^\dagger_j\hat{a}_i+\hat{a}^\dagger_i\hat{a}_j\right)+JS\sum_iz_i\hat{a}^\dagger_i\hat{a}_i+iDS\sum_{\langle ij\rangle}\eta_{ij}\left(\hat{a}^\dagger_j\hat{a}_i-\hat{a}^\dagger_i\hat{a}_j\right),
\end{equation}
with $N_l=\sum_iz_i$ the total number of links, where $z_i$ denotes the number of nearest neighbors of site $i$, i.e., $z_i = 4$ in the bulk, and $z_i\le 4$ on the edges. The spectrum of this model is represented in Fig.~\ref{fig:3bands}(b), where we indicate on the side the Chern number of each band.

\subsection{Chern marker measurement}
The protocol developed in the main text extends straightforwardly to systems with more than two bands. For the chiral perturbation defined in Eq.~\eqref{eq:chiral_pert}, the integrated differential excitation rate obtained from Fermi's golden rule reads
\begin{align}
        \int d\omega\left(\Gamma^{+}(\omega,t)-\Gamma^{-}(\omega,t)\right)&\underset{t\to+\infty}{\longrightarrow}-\frac{8\pi \epsilon^2}{\hbar^2}Im\left(\sum_{g\in LB}\left\langle \psi_0|g\right\rangle\langle g|\hat{x}\hat{Q}\hat{y}|g\rangle\left\langle g|\psi_0\right\rangle\right) \, ,
    \end{align}
where $\hat Q = \sum_{e \notin LB} \ket{e}\bra{e}$ projects onto all higher bands. 
The circularly polarized light excites all interband transitions, exactly as required:~the Chern number of the lowest band involves matrix elements coupling it to all other bands
\begin{equation}
    \mathcal{C}=4\pi\int\textrm{d}\Vec{k}\sum_{n>0}\Im\left(\langle u_0(\Vec{k})|\hat{x}|u_n(\Vec{k})\rangle\langle u_n(\Vec{k})|\hat{y}|u_0(\Vec{k})\rangle\right) \, ,
\end{equation}
where $u_m(\vec k)$ are Bloch states in momentum space. 
By covering all these transitions, the integrated chiral excitation rate for an initial state prepared in the lowest energy band therefore directly measures the Chern marker of the lowest band via Eq.~\eqref{eq:Chern_Mark_us}, regardless of the total number of bands. We verified this result numerically using the three-band Kagome lattice, and show the results in Fig.~\ref{fig:3bands}(c):~the Chern marker is well quantized in the bulk and takes large negative values near the edges, confirming the effectiveness of the protocol beyond the two-band case.

\putbib
\end{bibunit}
\end{document}